# Emotion Controlled Spectrum Mobility Scheme for Efficient Syntactic Interoperability In Cognitive Radio Based Internet of Vehicles

[1]Faisal Riaz, [2]Muaz A. Niazi
[1]Dept. Of Computing and Technology-Iqra University, Islamabad, Pakistan
[2]Dept. Of Computer Sciences-COMSATS, Islamabad, Pakistan
Email: [*]fazi_ajku@yahoo.com, muaz.niazi@gmail.com

*Abstract*— Blind spots are one of the causes of road accidents in the hilly and flat areas. These blind spot accidents can be decreased by establishing an Internet of Vehicles (IoV) using Vehicle-2-Vehicle (V2V) and Vehicle-2-Infrastrtructure (V2I) communication systems. But the problem with these IoV is that most of them are using DSRC or single Radio Access Technology (RAT) as a wireless technology, which has been proven to be failed for efficient communication between vehicles. Recently, Cognitive Radio (CR) based IoV have to be proven best wireless communication systems for vehicular networks. However, the spectrum mobility is a challenging task to keep CR based vehicular networks interoperable and has not been addressed sufficiently in existing research. In our previous research work, the Cognitive Radio Site (CR-Site) has been proposed as in-vehicle CR-device, which can be utilized to establish efficient IoV systems. However, the CR-Site is not capable to solve the spectrum mobility challenge in high speed IoV systems. In this paper, we have introduced the Emotions Inspired Cognitive Agent (EIC_Agent) based spectrum mobility mechanism in CR-Site and proposed a novel emotions controlled spectrum mobility scheme for efficient syntactic interoperability between vehicles. For this purpose, a probabilistic deterministic finite automaton using fear factor is proposed to perform efficient spectrum mobility using fuzzy logic. In addition, the quantitative computation of different fear intensity levels has been performed with the help of fuzzy logic. The system has been tested using active data from different GSM service providers on Mangla-Mirpur road. This is supplemented by extensive simulation experiments which validate the proposed scheme for CR based high-speed vehicular networks. The qualitative comparison with the existing-state-of the-art has proven the superiority of the proposed emotions controlled syntactic interoperable spectrum mobility scheme within cognitive radio based IoV systems.

# 1. Introduction

Blind spots are one of the causes of road accidents in the hilly and flat areas. According to the Mohanty and Gupta [1], accidents in the hilly area of Himachal Pradesh, India, are the 7[th] major reason of deaths. According to Talbot et al. [2], in the year of 2014, 113 light weight road commuters lost their lives due to the blind spot accidents in the United Kingdom. Riaz and Niazi [3] has reported that every year 300 citizens of United States lost their lives in blind spot accidents. These blind spot accidents can be decreased by establishing Internet of Vehicles (IoV), which help them to exchange their location and speed to avoid the collisisons before time. In this regard, Vehicle-2-Vehicle (V2V) and Vehicle-2-Infrastrtructure (V2I) communication mechanisms based IoV systems have been proven to be best.

Different type of IoV systems using V2V communication systems have been proposed. Xiang et al. [4] have proposed Dedicated Short Range Communication (DSRC) based V2V communication system to avoid the collisions between the vehicles. For this purpose, authors have equipped the vehicles with Global Positioning System (GPS), which help the vehicles to find their location, which they further transmit to the neighbouring vehicles through DSRC. A platoon based vehicle to vehicle communication system using IEEE 902.11p has been proposed by Bergenhem et al. [5] under the project of SARTRE. The aim of SARTRE project is to develop a technology by which vehicles could move in groups without any collision. Park et al. [6] have proposed an idea to use existing well-established cellular network based smartphone platform for vehicular networking applications. However, the problem with these IoV systems is that most of them are using DSRC or single Radio Access Technology (RAT) as a wireless technology, which has been proven to be failed for efficient communication between vehicles [7]. Hence, there is a need of new wireless technology, which help the vehicles to exchange their credentials more efficiently.

Cognitive Radio based V2V and V2I has to be proven best wireless communication system for IoV systems. Cognitive Radio-Enabled Optimal Channel (CROCS)-Hopping Sequence for Multi-Channel Vehicular Communications are presented by Chu et al.[8]. In CROCS scheme, cognitive radio concept is applied to safety and non-safety messages in Vehicular Adhoc Networks (VANET). Felice et al. [9] have proposed a cog-V2V framework for spectrum sensing, decision, and sharing in Cognitive Radio-Vehicular Ad hoc Networks (CR-VANET). In another research work, TV white spaces based V2V communication system using geolocation database and spectrum sensing has been proposed by Altintas et al. [10]. However, the spectrum mobility is very frequent in cognitive radio based IoV systems due to the highly mobile nature of vehicular networks.

In the existing literature different handover schemes for CR networks have been proposed. Kumar et al. [11] have proposed a spectrum handover scheme for optimal network selection in NEMO (Network Mobility) based CR vehicular networks using multiple attributes decision making. Furthermore, Kumar et al. [12] have proposed a game theoretic auction theory based spectrum handover scheme for CR vehicular networks. In another research work, Kumar et al. [13] have proposed a context-aware spectrum handover scheme with multiple attributes decision-making (MADM) method for CR vehicular networks. An efficient spectrum handover scheme for CR network using Fuzzy Logic and Neural Network has been proposed by potdar and patil [14]. The authors have considered Secondary User (SU) mobility, QoS, and priority as the main parameters for spectrum mobility. Nejatian et al. [15] have proposed a proactive channel switching scheme for Cognitive Radio- Mobile Adhoc Networks (CR-MANETs) considering SU mobility pattern. A fuzzy logic inspired channel selection procedure for efficient handover in cognitive radio based networks has been proposed by salgado et al. [16]. In another research work, Ahmed et al. [17] have proposed handover decision-making process using Primary User (PU) activity and Signal-To-Noise-Plus-Interference Ratio SINR as handover decision parameters along with optimal channel selection. However, the problem with [11-14] is that these handover schemes have been designed for only non-safety applications like voice and video services, which are delay tolerant. Whereas the spectrum mobility for safety applications with the minimal delay has not been addressed. Furthermore, these handover schemes have not addressed the two important phases of any spectrum handover, which are spectrum measurement and handover execution [18]. If we analyze the proactive handover schemes [15-17] then the problem with these handover schemes is that they have not been designed keeping in view the challenging nature of syntactic interoperability within high-speed cellular based vehicular networks. Here we need some sort of novel approach, which help in making the cognitive cycle of cognitive radio more efficient so that it makes spectrum mobility decisions efficiently in the high speed vehicular network environment. In this context, role of emotions in enhancing the cognitive functions of cognitive radio such as spectrum sensing, optimal whitespace selection and spectrum shifting can be explored.

Emotions enhance the efficiency of cognitive functions like perception, reasoning, and decision making. According to [19] emotions play a vital role in the flawless reasoning processes. Furthermore, It has been noted by Novianto and Williams [20] that emotions help in efficient decision making. Chen et al.[21] have stated that fear related emotions enhance the overall cognitive performance of the human beings during dangerous situations. According to Pourtois et al. [22], emotions help in enhancing the perception capabilities of human beings. However, this role of emotions has not been explored in the field of cognitive radio in general and in the spectrum handover for efficient syntactic interoperability in vehicular networks in specific.

In our previous research work, we proposed the outlines of Cognitive Radio Site (CR-Site) that acts as in-vehicle gadget for efficient V2V communication [23]. The CR-Site contains following functions (i) spectrum sensing (ii) spectrum optimisation and (iii) spectrum mobility. In [23] we have given the details of CR-site along its spectrum optimisation module, which helps in selecting optimal white space with the help of a Memory Enabled Genetic Algorithm (MEGA). Though the spectrum mobility module exists in CR-Site but its details were not provided in [23]. In the current research, the role of fear emotion has been explored to propose complete three phases, handover measurement & initiation, handover decision and handover execution, spectrum mobility protocol as compared to [11-14] for the CR-Site in the high speed vehicular networks. To the best of our knowledge, the role of emotions is going to be explored first time in this context. In a conclusion, we are going to expound novel fear based spectrum mobility module of CR-Site. In the perspective of the big picture, four main contributions are made in this paper.

- First, the role of emotions has been explored using Exploratory Agent Based Modelling level of CABC framework to enhance the efficiency of spectrum mobility feature, for efficient syntactic interoperability between vehicles, of the CR - Site. For this purpose, a novel architecture of EIC_Agent has been proposed.

- Second, a probabilistic deterministic finite automaton (PDFA) is designed, which is acting as a brain for the spectrum mobility module of EIC_Agent. PDFA helps the spectrum mobility module to answer "When" and "Where" spectrum mobility could be performed.

- Third, quantitative measurement of different fear intensity levels using fuzzy logic has been performed.

- Fourth, the validation of the proposed EIC_Agent based spectrum mobility scheme using Validated Agent Based Modelling level of CABC framework has been performed.

The remaining paper is arranged as follows. Section 2 is discussing the proposed methodology. Overview of the existing CR-Site architecture has been presented in section 3. Section 4 discusses proposed EIC_Agent architecture and its role in extending the architecture of the CR-Site. The subsection of section 4 is reporting proposed fear inspired PDFA. Section 5 presents the designed experiments. Section 6 is elucidating the agent based simulation. Section 7 is the results and discussion section. The qualitative comparison with the existing-state-of-the art has been made in section 8. Section 9 is concluding the paper.

## 2. Proposed Methodology

The big picture of the overall methodology has been presented in figure 1. Before proposing EIC_Agent based spectrum mobility scheme, a field survey was conducted in Mirpur Azad Kashmir to build the spatial database consisting of homogeneous multi-RAT's and heterogeneous service providers' signal strengths. A block diagram is pictorially presenting an overview of the overall methodology. The survey started from the F.2 road near the main Mosque (starting point) all the way to Mangla (ending point) of total 8 km distance. The F2-Mangla road is the one where cellular users have a bad signal strength problem. The purpose of the survey was to measure the GPS points where the different GSM service providers (SP1, SP2, SP3, and SP4) signals are absent or having bad signal strength. The reason for choosing only one spectrum band, i.e. 1800 MHz is to decrease the packet latency time due to the operating frequency reconfiguration of the RF-front end of CR-device installed within AV. The gathered GPS locations were then mapped on Google map to demonstrate the absence of excellent, good and bad quality signals shown in figure 2. To conduct the survey two main kinds of hardware was used. Single antenna GPS was used to measure the (latitude & longitude) and Cell Track software was used to measure the signal strength of different GSM operators on the F2-Mangle road. Table 1 presents the readings of the survey, including the (latitude, longitude) along with the different signal strengths of three major GSM providers (SP1, SP2 & SP3). These readings act as a pre-database for a vehicle moving on F2-Mangla road to evade on-going transmission interrupts by performing early spectrum decision using MEGA detailed in [7] and hence in time spectrum mobility. The readings of table 1 are then mapped on a Google map of the Mirpur city as shown in figure 6 to use it for simulation and validation of our proposed FISMS algorithm.

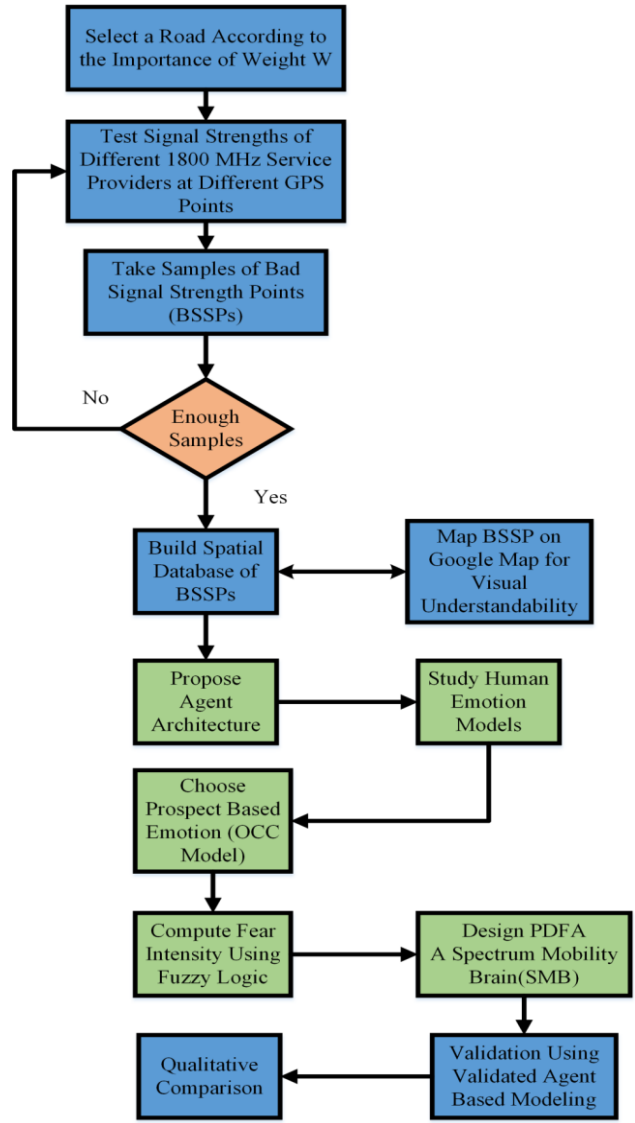

Fig. 1 Big Picture of Overall Methodology

## 3. Overview of the Existing Cognitive Radio-Site

The existing cognitive site architecture, shown in Fig. 2, consists of four main modules, those are:

- Scanner: Scans the available Radio Access Technologies (RATs) within the range of vehicle movement.

- Sensor: Senses the white spaces if opportunistic access mode is set else simply shift on another available spectrum band.

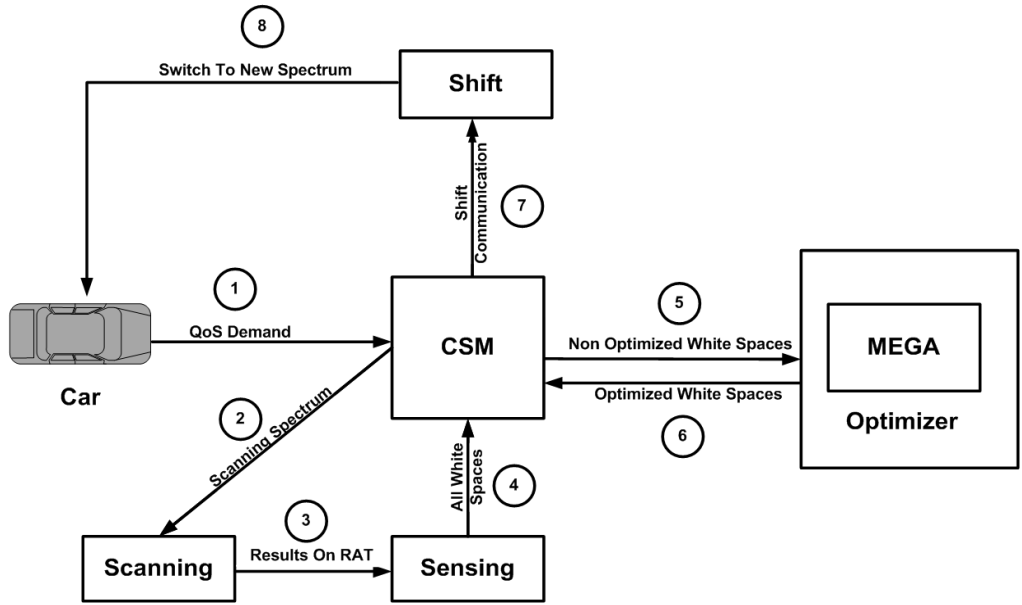

Fig. 2 Existing CR-Site Architecture [23]

- Optimizer: Finds the optimised white spaces in order of priority using efficient spectrum decision using MEGA module according to the vehicle QOS request.
- Analyser: Continuously analyses performance of allotted white-space to perform

A cognitive system monitor (CSM), acting as the brain of the CR-Site, coordinates the four modules. To begin with, a vehicle sends a call request to CSM in order to get access to a suitable channel. The QoS requirements of the vehicle are also included in the request. Upon receiving this request, CSM asks the scanning module to scan all RATs in search of a set of available channels. The scanning module generates a list of available RATs and their respective channels. This list is passed on to the sensing module. A variety of sensing methods such as cooperative sensing [24] and cyclo-stationary techniques [25] have been proposed. The main task of the sensing module is to detect the white spaces within the available bands. The sensing module generates and passes on this list to CSM, which modifies the pool of available white spaces. The sensing module uses the dual radio sensing architecture such that one radio chain is dedicated for data transmission and reception, while the other chain is dedicated for spectrum monitoring [26]. The sensing module also keeps sensing the presence of a primary user. Once a PU is detected, white spaces are computed afresh. Once the pool of available white spaces is ready, CSM employs Memory Enabled Genetic Algorithm (MEGA) to determine the optimum white space that should be used by a vehicle. The optimisation performed by MEGA also takes into account the QoS requirements passed on by the vehicle.

However, the problem with this existing CR-Site is that though the Shift module is the part of CR-Site the details of its function have not been presented in [7]. Keeping in view this problem, we have proposed human emotions inspired proactive spectrum mobility component in CR-Site instead of simple Shift module. It is important to mention that we have not made changes in existing functions of CR-Site and the proposed spectrum mobility module is working in line with other modules of existing CR-Site. To achieve the proactivity, human emotion inspired agent has been proposed and its details are given in next section.

# 4. Proposition of Emotions Inspired Cognitive Agent Architecture to Extend the Spectrum Mobility Capabilities of CR-Site.

The proposed EIC_Agent architecture and extended version of the CR-Site for efficient spectrum mobility has been presented in figure 3. The proposed agent consists of four main modules. Their description is given as under.

1. **Signal Strength Measuring Module:** This module helps the agent in measuring the signal strength of different white spaces available in white space pool maintained by a CSM module of CR-Site.
2. **GPS Based BSSPS Database Module:** This module contains the information regarding all BSSPS.
3. **Fear Generation using OCC Module:** This module helps the agent to feel the fear according to the mechanism defined by OCC model.
4. **CSM Control Rules Selection Module:** This module helps the agent to control the functions of CSM under the following rules.

    i. **IF** *Fear_Intenisty* is Very Low or Low

    **THEN** Keep Communicating Using Current_Service

    ii. **IF** *Fear_Intenisty* is between High_Low and Low_Medium

    **THEN** Initiate Spectrum Sensing Instruction to the CSM

    iii. **IF** *Fear_Intenisty* is between High_Medium and Low_High

    **THEN** Initiate Spectrum Optimizer Instruction to the CSM

    iv. **IF** *Fear_Intenisty* is Medium_High

    **THEN** Initiate Spectrum Handover Instruction to the CSM

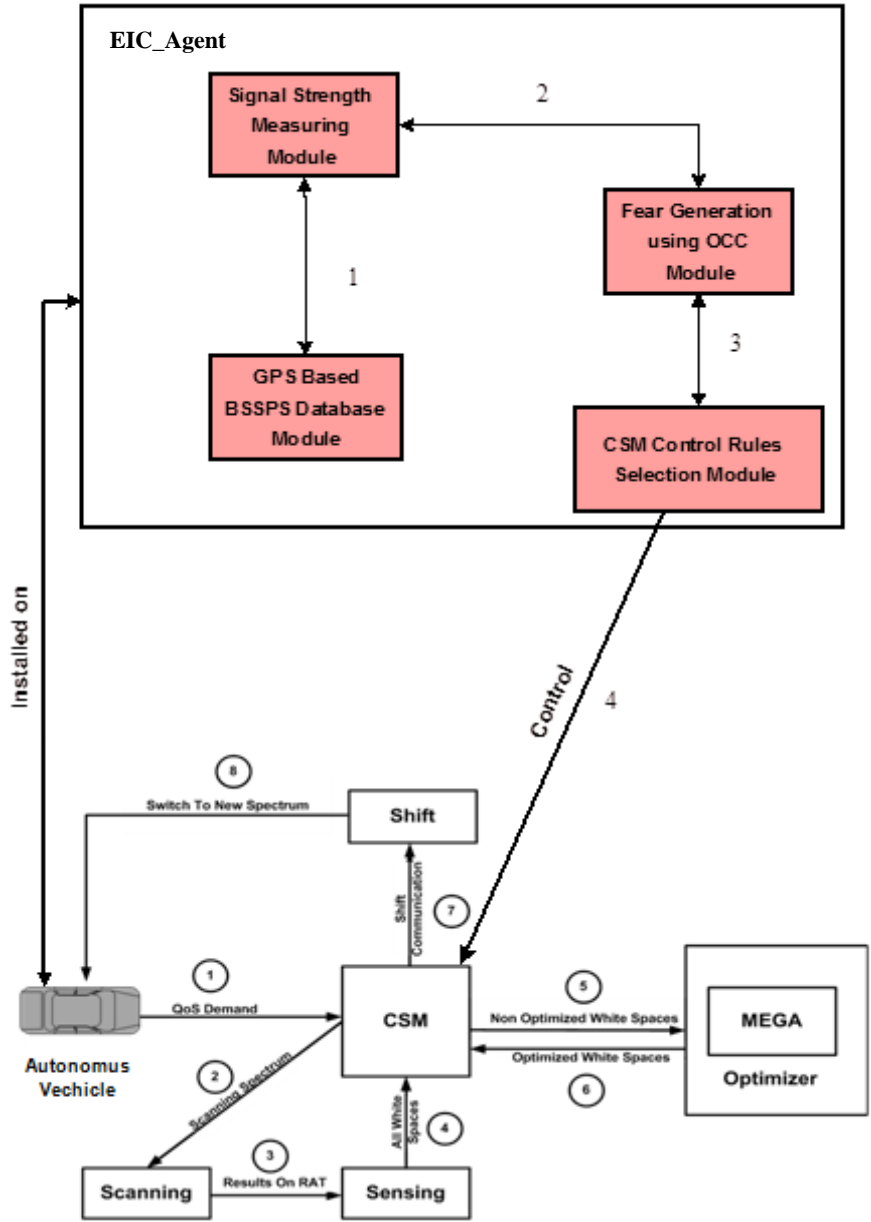

Fig. 3 EIC_Agent extended CR-Site

It has been assumed that the Both CR-Site and EIC_Agent are installed on AV. From the figure 3, it can be seen that EIC_Agent control the functionality of CSM according to the rules defined in its CSM Control Rules Selection Module. The next subsections help the readers to understand the detailed description of the proposed EIC_Agent architecture, its modules and step by step functionality.

## A. Building GPS Based BSSPS Database

As discussed in methodology section a detailed survey regarding the different signal strengths of different service providers has been performed on the Mirpur-Mangla road to build the database of BSSPs. Table 1 is presenting BSSPs along their GPS coordinates and signal strengths. For further clarity, thesome of the BSSPs points among the collected samples have been maped to the Google map as shown in figure 4.

TABLE 1 SIGNAL STRENGTHS OF DIFFERENT GSM SERVICE PROVIDERS ON DIFFERENT GPS LOCATIONS OF F2-MANGLA ROAD

| Measuring Points | GPS Coordinates | | SP1 Signal Strength (dBm) | SP2 Signal Strength (dBm) | SP3 Signal Strength (dBm) |
|---|---|---|---|---|---|
| | Latitude | Longitude | | | |
| A | 33.144552 | 73.745719 | -100 | -90 | -80 |
| B | 33.144449 | 73.745606 | -60 | -70 | -50 |
| C | 33.144377 | 73.745550 | -50 | -60 | -50 |
| D | 33.144323 | 73.745475 | -65 | -50 | -70 |
| E | 33.144235 | 73.745386 | -80 | -30 | -50 |
| F | 33.144163 | 73.745300 | -45 | -50 | -75 |
| G | 33.144132 | 73.74520 | -65 | -95 | -85 |
| H | 33.144076 | 73.745064 | -75 | -75 | -100 |
| I | 33.144051 | 73.744965 | -90 | -50 | -40 |
| J | 33.144029 | 73.744850 | -70 | -70 | -70 |
| K | 33.144043 | 73.744742 | -50 | -60 | -65 |
| L | 33.124631 | 73.665952 | -85 | -91 | -81 |
| M | 33.122716 | 73.665782 | -83 | -96 | -100 |
| N | 33.118999 | 73.663482 | -81 | -100 | -78 |

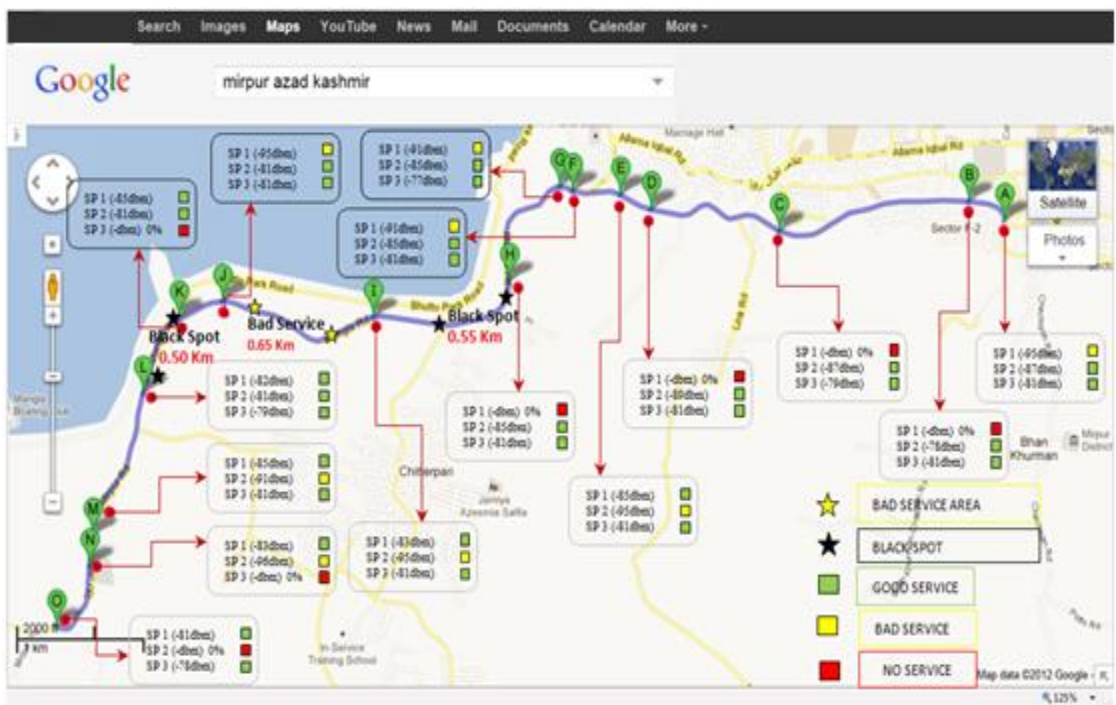

Fig. 4 Google map mapped with the BSSPs

## B. Proposed Fear Generation Mechanism of EIC_Agent

It is important to note that emotions are the product of the appraisal of the contemporary state [27]. Humans use two mechanisms for the assembly of their judgment named intuition (rapid and emotional) and reasoning (deliberate and precise). However, both of the approaches rely on the appraisal theory regarding the generation of fear. The fearful situation can be tackled by automatic or deliberative mechanisms. In our work vehicles are tackling fear situation by employing reasoning process (e.g. Due to the bad signal strength, I will lose white space). The intensity of the fear experienced by vehicle can also be effected as a feedback function of this reasoning process (e.g. If I lose the white space, I will be failed to exchange my position with other vehicles and if I will not exchange my position I will collide with any vehicle). The vehicle response to this situation will be conscious (e.g. I have to perform spectrum mobility before reaching the bad signal point).

The fear can be defined as an adaptive retort to the frightening condition [28]. However, these frightening conditions can be innate or learned. This section is revealing how the treacherous state will be detected (appraisal of fear) by the spectrum mobility module of CR-Site. The notion is analogous to a person who knows about the hazards on a jogging track due to his daily routine. Whenever he approaches anyone of that hazardous point he becomes alerted and takes encountering measures like slowing down the speed or taking a long jump to avoid the hazard.

In the light of above discussion, we have proposed a Fear-Inspired Spectrum Mobility Scheme (FISMS) which is incorporating affective computing for an efficient spectrum handover. The efficiency of spectrum mobility scheme is achieved in terms of in-time white space allocation to the vehicles. For this purpose, we have considered the best-categorized emotion model, Orton, Collins (OCC) model [29]. OCC model divides emotions into three main categories i.e. Event-based emotions, Attribution emotions and Attraction emotions. Further Prospect-based emotions are the subcategory existing under the hierarchy of event based emotions and well address our problem. Prospect-based emotions can be defined as the emotions experienced in response to predictable or alleged events [29] i.e. in our case vehicle mobility module of CR-Site experience fear in response to the expected event of passing through the bad signal point and losing data communication channel.

Our work is relying on learned releaser of fear which is here the bad signal point. The system proposed in this paper is relying on prior information of bad signal points (F2-Mangla road) as appraisal mechanism of fear. The intensity of fear is controlled with the help of distances between these bad signal points. Bad signal points are acting as dangerous states and causes fear generation i.e. vehicle will feel fear when it will travel towards these points and the intensity of fear will vary according to the less or more distance. As the distance between bad signal strength point and vehicle will decrease the fear will appraise and vice versa. For the appraisal of fear and proactive spectrum mobility scheme, a PDFA is proposed in the C section. In addition, we have introduced affective computing in our work to enhance the efficiency of our proposed novel spectrum handover schemes in the context of pure proactive spectrum handover methodology.

*Computing the Emotions*

Human emotions are measured using qualitative terms like "I feel average fear" or "He seems little tired". Numerical values could not be used to measure the intensity of human emotions. Following this fact, it is more natural to associate the qualitative terms with the fear. Fear can be expressed using high intensity, average intensity, low intensity, very low intensity etc. These mentioned terms can be measured in terms of some relevant degree using fuzzy logic as elucidated in [30]. Fuzzy logic is successfully evaluated for diverse purposes, like control devices [31], pattern recognition [32], social complexity [33] and human emotions emulation and simulation in intelligent agents [34]. In [35] artificial human emotions are generated using fuzzy logic to implement the emotions inspired intelligent agents. In [35] an agent adapts different emotions according to the different events and observations. In the light of research work presented in [35] fuzzy logic is found as a suitable tool to implement the fear inspired spectrum mobility scheme. The EIC_Agent measures distance, signal strength and feels fear accordingly. According to the fear rate, EIC_Agent makes in-time decision to perform the in-time spectrum mobility before the loss of underused whitespace. In the next section, experiment1 has been performed, which help in computing the quantitative values of emotion using fuzzy logic.

## C. Proposed Spectrum Mobility Mechanism of EIC_Agent

In the perspective of the big picture, a probabilistic deterministic finite automaton (PDFA), figure 5, is designed which is acting as a brain for the spectrum mobility module of EIC_Agent extended CR-Site. PDFA helps the CR-Site to answer "When" and "Where" spectrum mobility in case of bad RSS could be performed. It is important to mention that the current work is not addressing the spectrum mobility due to the Primary User (PU) activation.

*Deterministic Finite Automaton: A spectrum mobility Brain*

A deterministic finite automaton (DFA) is designed which acts as a brain for the spectrum mobility module of CR-Site. The rate of fear acts as a decision entity for the state-state transition. The transition table of proposed PDFA using different fear intensities is shown in table 2. PDFA can be defined properly as a 5-tuple $K = (Q, \mathcal{W}, \delta, \phi, r_0)$. Here Q represents a finite set of states which in our case are $Q = \{1, 1a, 1b, 2, 2a, 2b, 3, 3a, 3b\}$ where state 1 is representing SP1, 2 is SP2 and 3 is SP3, $\mathcal{W}$ is representing set of finite alphabets of observable symbols which in our case are $\mathcal{W} = \{S, M, I, C\}$, where S is representing transition to the self-state if fear intensity is $0.2 < \phi < 0.3$, M is representing a transition to next or previous state depending on the fear intensity in the range of $0.4 < \phi < 0.6$, I representing for initiating the optimizer module to compute the optimized white space if fear intensity lies in the range of $0.6 < \phi \leq 0.8$ and C to perform the spectrum mobility when fear intensity approaches $\phi > 0.8$, $\delta = Q * \mathcal{W} \rightarrow Q$ is the transition function, $\phi = Q * \mathcal{W} \rightarrow [0, 1]$ is the probability of next symbol from $\mathcal{W}$ and is used to measure the rate of fear in our case, lower the probability lower the fear and vice versa. Lastly $r_0$ is the initial state, which can be any one among Q in our case and it depends on the current GPS location of the vehicle and information of the available GSM service provider on current GPS location. That is why it can be seen in figure 7 that we have designated more than one initial state in proposed PDFA. It has been assumed that the AV has spatial pre-database of multi GSM service provider's signal strength stored in the on-board unit (OBU) of EIC_Agent is acting as the main module to manipulate this database. When the AV starts travelling, then EEC_Agent computes the current latitude and longitude of the vehicle by initiating Garmin-Terex legend GPS hardware. Current available RAT information on current GPS location is loaded from the pre-database and it helps to choose the initial state among Q = {1, 1a, 1b, 2, 2a, 2b, 3, 3a, 3b}. For example the current available GSM operator on computed GPS coordinates is SP1 then the state 1 of PDFA is the entry point. The GPS computing module of vehicle checks twice in a second the GPS coordinates of vehicle and match it with pre-database information in the state 1. The function $\delta = Q * \mathcal{W} \rightarrow Q$ will get its value according to the probability computed by function, $\phi = Q * \mathcal{W} \rightarrow [0,1]$ on the basis of $D = \gamma_{CGL} - \beta_{DZGL}$, where $\gamma_{CGL}$ and $\beta_{TGL}$ are the variables holding the vehicle's current and incoming danger zone GPS locations respectively. On the basis of computed probability the transition from 1 to 1a, 1b, 2, 2 to 2a, 2b and 3 to 3a, 3b will be occurred as shown in state transition table 2 for real time spectrum mobility before reaching the $\beta_{TGL}$

TABLE 2 STATE TRANSITION TABLE OF PDFA USING FEAR FACTOR

| State | Same (S) | Move (M) | Initiate Spectrum Optimisation (I) | Spectrum Handover (C) |
|---|---|---|---|---|
| 1 | 0.2 < f < 0.3 (1) | 0.4 < f < 0.6 (1a) | --- | --- |
| 1a | 0.4 < f< 0.6 (1a) | 0.2 < f < 0.3 (1) | 0.6 < f <=0.8 (1b) | --- |
| 1b | 0.6 < f <=0.8 (1b) | 0.4 < f < 0.6 (1a) | --- | *f > 0.8 (Suitable and available service) |
| 2 | 0.2 < f < 0.3 (2) | 0.4 < f < 0.6 (2a) | --- | --- |
| 2a | 0.4 < f < 0.6 (2a) | 0.2 < f < 0.3 (2) | 0.6 < f <=0.8 (2b) | --- |
| 2b | 0.6 < f <=0.8 (2b) | 0.4 < f < 0.6 (2a) | --- | *f > 0.8 (Suitable and available service) |
| 3 | 0.2 < f < 0.3 (3) | 0.4 < f < 0.6 (3a) | --- | --- |
| 3a | 0.4 < f < 0.6 (3a) | 0.2 < f < 0.3 (3) | 0.6 < f <=0.8 (3b) | --- |
| 3b | 0.6 < f <=0.8 (3b) | 0.4 < f < 0.6 (3a) | --- | *f > 0.8 (Suitable and available service) |

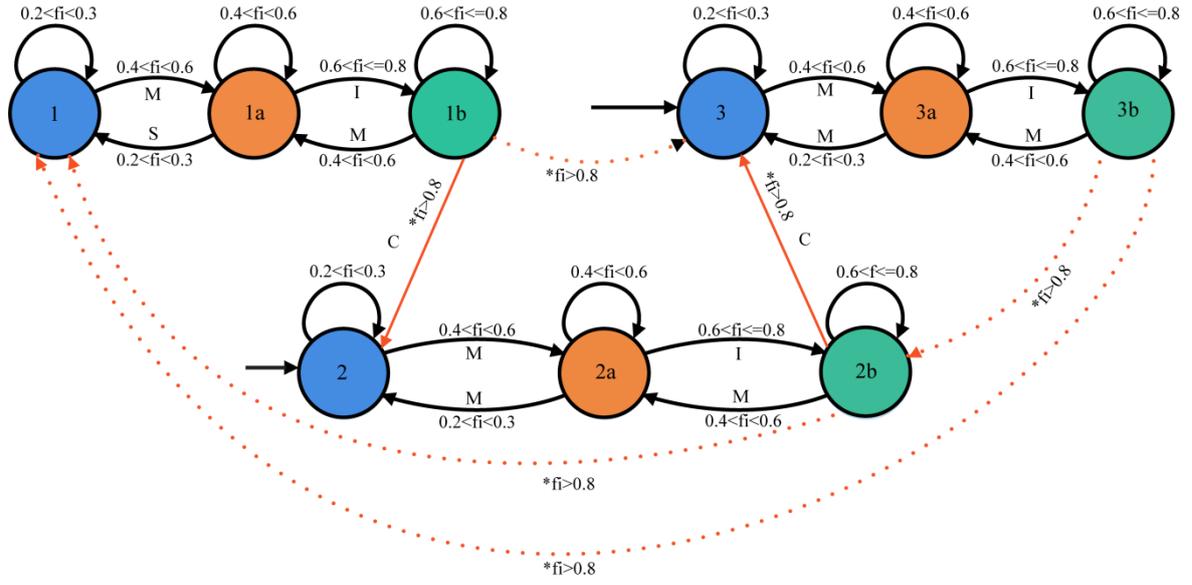

Fig. 5 Spectrum Mobility Mechanism of EIC_Agent Using Fear based PDFA

# 5. Experiments

In this section, two main types of experiments have been performed. The first one experiment has been performed to compute the quantitative values of Fear using fuzzy logic. Then the second type of experiment helps in validating the different features and functionalities of proposed EIC_Agent.

## 5.1 Experiment 1: Quantitative Computation of Fear Using Fuzzy Logic

To compute the different intensity levels of fear, we have built a Mamdani fuzzy inference system, which uses the traceability algorithm defined in [29].

   If Prospect (v, e, t) and Undesirable (v, e, t) < 0

      Then set Fear-Potential (v, e, t) = $f_f$ [|Desire (v, e, t) |, Likelihood (v, e, t), Ig (v, e, t)]

   If Fear-Potential (v, e, t) > Fear-Threshold (v, t)

      Then set Fear-Intensity (v, e, t) = Fear-Potential (v, e, t) - Fear-Threshold (v, t)

      Else set Fear-Intensity (v, e, t) =0

In order to compute the Fear-Potential of the computation traceability algorithm, we have to calculate the values of Likelihood, Desirability, and Intensity of a global variable.

### 5.1.1 Computing Likelihood

Two input variables distance and signal strength are defined because the shifter module of the CR-Site uses distance and signal strength as decision variables to feel the fear in the case of possible data communication loss. In fuzzy logic, linguistic variables are used to express the ideas qualitatively. The pictorial description of fuzzy vocabulary for the Distance, Signal Strength and Fear can be seen in figures 6a, 6b and 6c. During very low-intensity or low-intensity fear state vehicle transmits its data using the same white space. As the shifter module of the vehicle feels average fear-intensity it initiates the spectrum optimizer module of the CR-Site to find the optimised white space according to the users' QoS parameters as discussed in details in [7] and performs the in-time spectrum mobility on feeling very high-intensity fear.

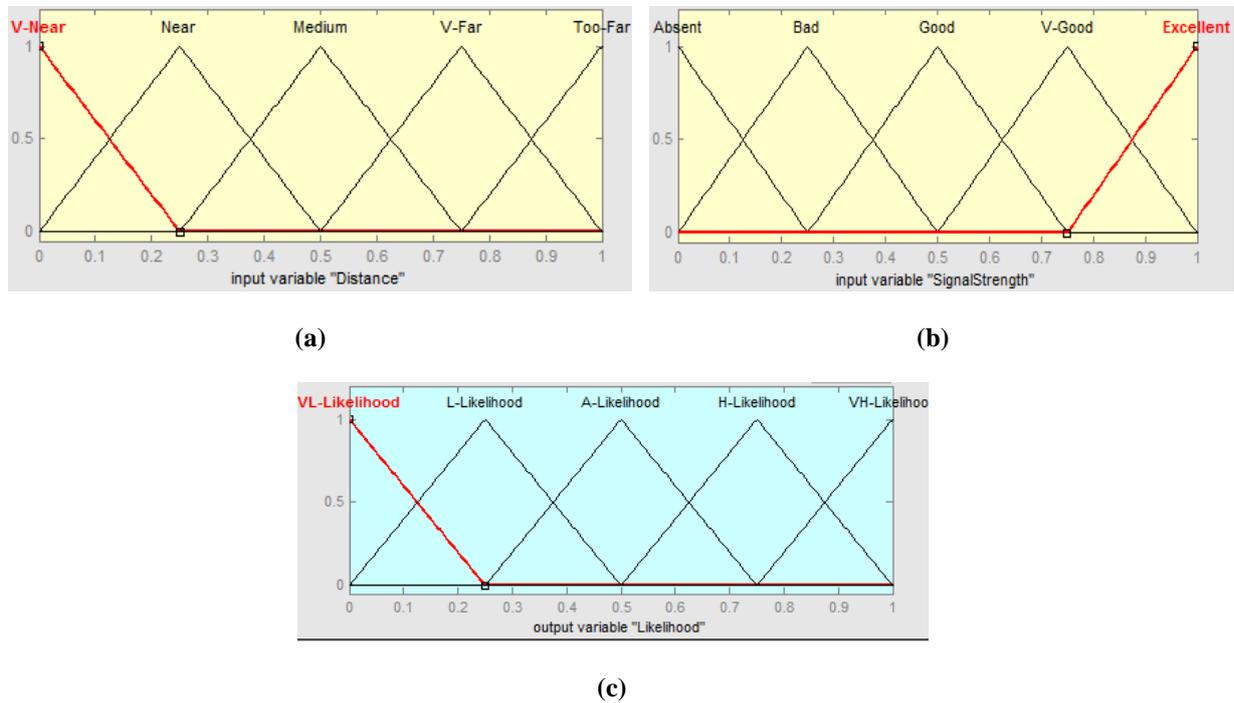

Figure. 6 Membership functions. (a) Distance variable, (b) Signal Strength variable, (c) Likelihood variable

In the simulation, distance variable consists of {V-Near, Near, Medium, V-Far, Too-Far} and its range is set between (0-1) linguistic words as shown in figure 6a. Signal strength variable consists of {Absent, Bad, Good, V-Good, Excellent} linguistic words and its range are set as (0-1) as shown in figure 6b. The output variable Likelihood consists of {VL-Likelihood, L-Likelihood, A-Likelihood, H-Likelihood, VH-Likelihood} and its range is set as (0-1) as shown in figure 6c. The main simulation screen of Computing Likelihood using a fuzzy logic toolbox is shown in figure 7. It can be seen that distance and SignalStrength are two input variables and Likelihood is an output variable.

### 5.1.2 Computing Undesirability

The second variable used to compute the fear intensity level of AV regarding signal loss problem is Undesirability. The Undesirability variable in our case has been taken in the context of the importance of communication between AVs. For example, if the road terrain is hilly and the vision sensors of AV are not able to detect hidden AVs ahead, then the Undesirability of signal loss is very high in this case. We have computed Undesirability by computing the importance of communication and signal strength variables. Figure 8 is presenting the main simulation screen of Computing Undesirability using a fuzzy logic toolbox.

### 5.1.3 Intensity of Global Variable

The Ig variable actually presents the human capability of vivid interpretation of prospect event. For example, a person is flying in an aeroplane and imagining an air crash, then the high Sense of Reality (SOR) and high VirtualtimeProximity (VTP) will contribute to the high fear intensity level. It is important to mention that SOR and VTP are mental level variables and in our case, these variables will be computed automatically as the AV will start travelling. For the sake of brevity, we are considering their high values only. In our case, sub-variable SOR defines the mental level interpretation of the importance of safety, and VTP is the mental level interpretation of the virtual time proximity of event occurrence, loss of signal strength in our case. Figure 9 is presenting the main simulation screen of Computing Ig variable using the fuzzy logic toolbox.

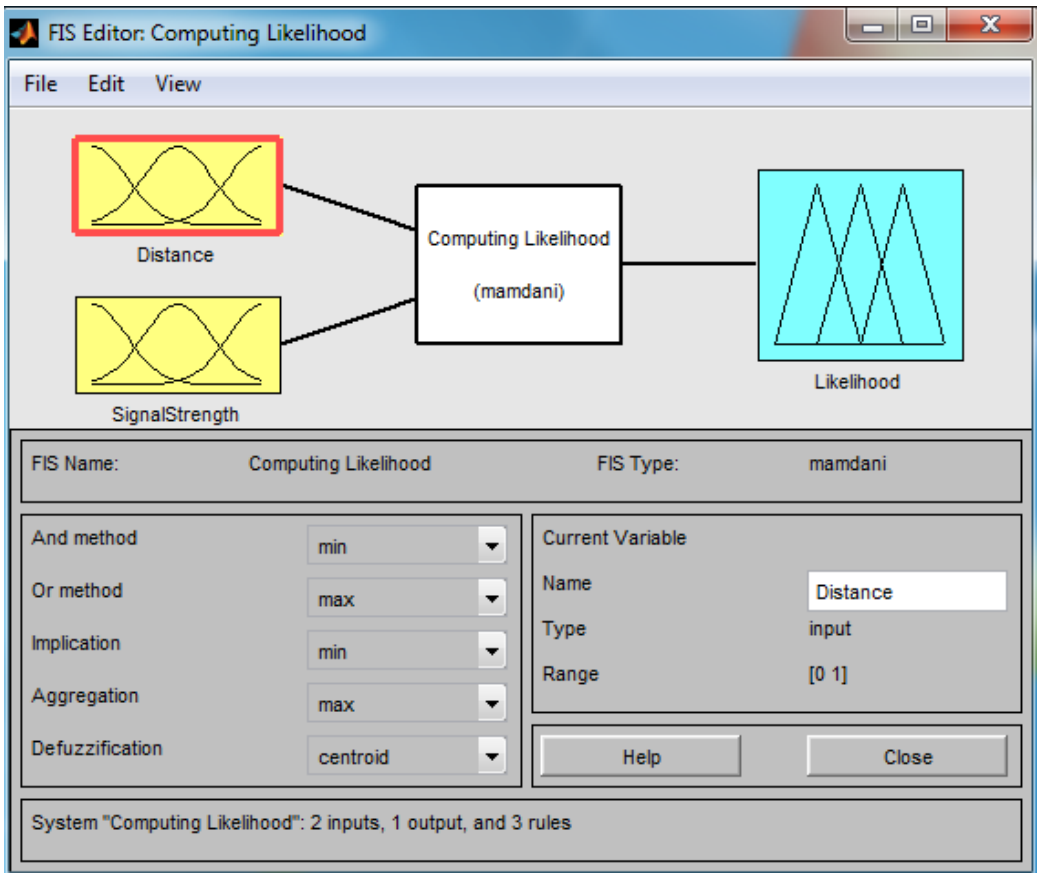

Fig 7 Main simulation screen of Fear generation system

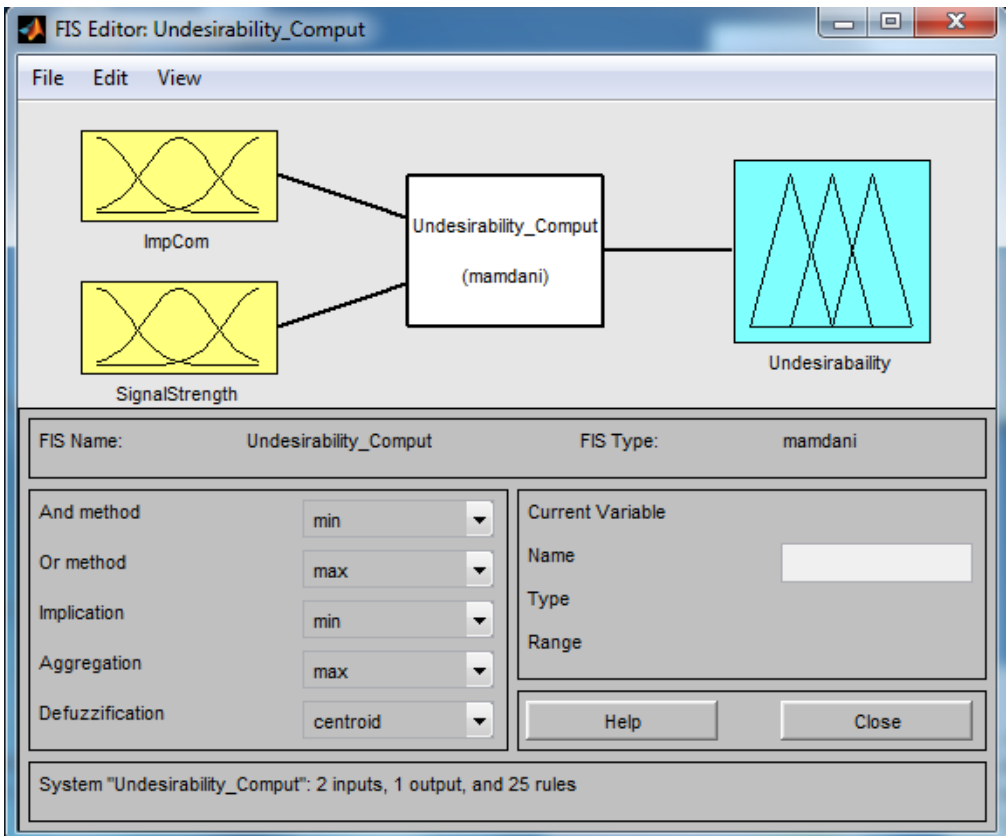

Fig. 8 Main simulation screen for Undesirability computation

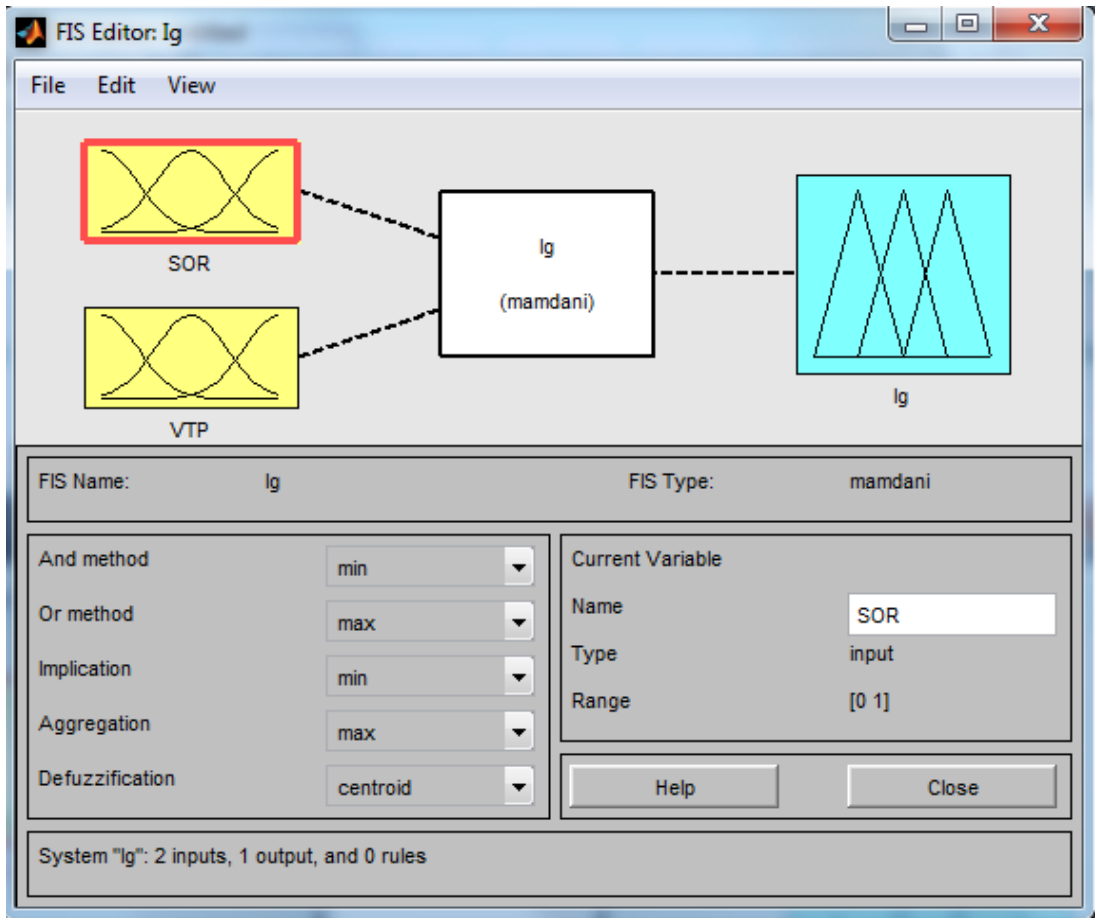

Fig. 9 Main simulation screen for Ig computation

## 5.2 Experiment 2: Validation of the Proposed EIC_Agent Functions and Features using VOMAS Agent-Based Simulation

The purpose OF experiment 2 is validating the features and functionalities of EIC_Agent. For this purpose, we have proposed the validation model, and then UML design of agent-based simulation has been proposed. In the last, the simulation details have been provided.

### 5.2.1 Validation Model

We have proposed VOMAS agent-based simulation validation methodology. First, we have elucidated the VOMAS agent design, and then we have raised the validation question. In the last, three invariants have been defined which will act as filters to validate the agent-inspired simulation of EIC_Agent based spectrum mobility scheme.

***VOMAS Agent Design***

In our simulation model, we have designated Autonomous Vehicle as a VOMAS agent. The VOMAS agent computes the changes in AV's fear, according to the OCC model defined fear related variables and perform spectrum mobility scheme efficiently.

*Validation Question*

Our validation questions can thus be defined as.

Question1: How can we validate that the enhanced emotion enabled cognitive agent is feeling different levels of fear according to the distance left between its current GPS point and the incoming Bad Signal Point (BSSP). Furthermore, how we can validate that the proposed PDFA based spectrum mobility brain is performing spectrum mobility according to the defined functionality.

*Invariants*

Three invariants are defined according to the guidelines given by [36] to find out the answers to validation question.

Invariant1

If the pre-condition that **"Distance between the EIC_Agent extended CR_Site installed AV and BSSP is decreasing"** is true, then the variation in the distance would result in a post-condition of **"Fear intensity of an EIC_Agent extended CR_Site installed AV is increasing accordingly"**.

Invariant2

If the pre-condition that **"The future signal strength of whitespace, in-use, is getting bad "** is true, then the variation in the signal strength would result in a post-condition of **"The EIC_Agent extended CR_Site will select new white space with good signal strength"**.

Invariant3

If the pre-condition that **"Distance between the EIC_Agent extended CR_Site installed AV and BSSP is decreasing and future signal strength of whitespace, in-use, is bad as well"** is true, then the variation in the distance and signal strength would result in a post-condition of **"The EIC_Agent extended CR_Site will complete spectrum mobility task (Sensing + Whitespace Optimization + TCP Connection Setup) before reaching the BSSP"**

*Tests Design for Invariant 1 and 2 and 3*

To find out the answers of Invariant1, 2 and 3, the power of BehaviorSpace has been utilized. BehaviorSpace is a special feature of NetLogo, which helps in performing the extensive testing. The test configuration is given as follows.

(1) EIC_Agent installed AV starts travelling from the GPS coordinates (33.144552, 73.745719) and ends its journey at GPS coordinates (33.144029, 73.744850). These GPS values have been taken from the table 1.

# 6 Simulation Environment

To test the performance of the proposed spectrum mobility scheme, we have utilized NetLogo 5.2. The NetLogo platform has been utilised because of its flexibility in implementing the agent-based systems. In our simulation, we have considered the AV travelling on the Mangla road as an agent, which seeks different spectrum slots to keep in communication with other vehicles for safe travel. The main simulation screen of the proposed spectrum mobility scheme is presented in the figure 13. The simulation environment has been designed keeping in mind the all collected real field data in the survey of Mirpur-Mangla road. It can be seen that the road in simulation contains all the information regarding BSSPs in the shape of GPS coordinates. The upper panel of the simulation contains the functions of the CR-Site such as Sensing, Optimal white space selection using MEGA and TCP setup connection time. The right panel helps in controlling the functions of autonomous vehicle. It includes the speed of the car, car start position, car stop position and random start position simulation elements. The lower panel helps in adjusting the values of fear related variables.

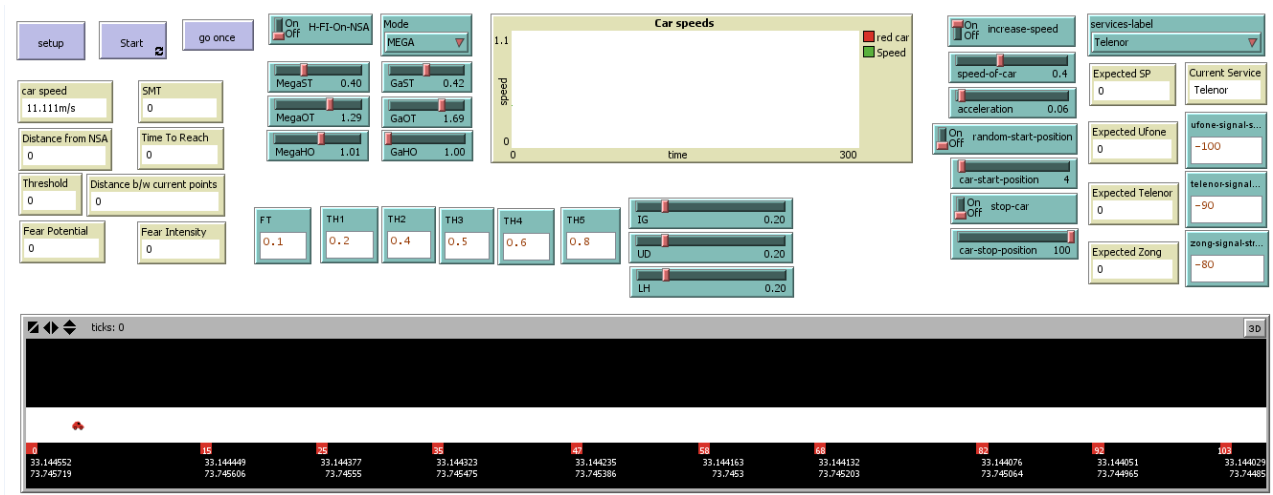

Fig. 13 Main Simulation Screen of Social Norms and Emotions inspired artificial society of AVs

## *6.1 Simulation Parameters*

The simulation parameters utilised in testing the proposed affective computing inspired spectrum mobility scheme has been presented in a table 3 with sufficient description.

TABLE 3
SIMULATION PARAMETERS AND THEIR DESCRIPTION

| Simulation General Parameters | Range | Description |
|---|---|---|
| Speed of Car | {0 -1 with increment of 0.1} | This slider helps in defining the travelling speed of AV. |
| Acceleration | {0.01-0.1 with increment of 0.01} | This slider helps in defining the acceleration rate of the AV. |

| Car-Start-Position | {1-100} | This slider helps in defining the starting position of AV within the simulation world. Actually, these are different GPS locations presented in table 2. |
|---|---|---|
| Car-Stop-Position | {2-100} | This slider helps in defining the stoping position of AV within the simulation world. Hence, using Car-Start-Position and Car-Stop-Position sliders, the performance of proposed scheme can be tested within different points of simulation world. Actually, these are different GPS locations presented in table 2. |
| Services-Label | {Telenor, Ufone, Mobilink, Zong} | This slider helps in defining the initial service provider, Like Telenor, Ufone or Zong. |
| Mode | {MEGA} | This slider helps in choosing the White space optimisation algorithm like MEGA. |
| CRST( Cognitive Radio Sensing Time, MEGAOT (White space optimisation time taken by MEGA, HOT (Handover Time) | Sensing time between {50ms-200ms} and these sensing time values have been taken from IEEE 802.22 standard [37]. The values of MEGA optimisation time have been taken from [7]. The TCP connection setup delay in a typical GPRS network has been considered between {1s-5s} as mentioned in [38] | These three options help in defining different sensing times, optimisation times, and handover times respectively in the case of MEGA as white space optimisation algorithm. |
| TH1, TH2,TH3,TH4,TH5 | {0.1-1} | These are the thresholds, which help in taking different decisions in the processes of spectrum mobility. For example, TH1 defines the fear intensity level, when the vehicle has to stay in the state 1of figure 7. For more details, table 2 can be consulted regarding different values and corresponding actions of thresholds TH2, TH3, TH4, and TH5. |
| **Prospect Based Emotion i.e. Fear Generation Parameters** | **Range** | **Description** |

| Likelihood | {0–1}; with increment of 0.1. For simulation purpose the values have been taken from table 5. | This slider helps in defining the likelihood of accident perceived by AV |
| --- | --- | --- |
| Desirability | {0-1; with increment of 0.1}. For simulation purpose the values have been taken from table 4. | This slider helps in defining the current desirability value of AV. |
| Ig | {0-1; with increment of 0.1}. For simulation purpose the values have been taken from table 3. | This slider helps in defining the current Ig value of AV. |

# 7  Results and Discussion

This section elucidates the results of Experiment 1 and Experiment 2 along the detailed discussion.

## 7.1  Experiment 1

Table 4 shows the quantitative values of undesirability from very low (VL) to very high (VH). The terms VLD, LD, MD, HD, and VHD represent very low desirability, low desirability, medium desirability, high desirability and very high desirability respectively. If the agent has a value between 0-0.24 for its undesirability of an event, then it can be interpreted as the very low undesirability. However, from an abstract analysis, it can be noted that due to the fuzzy nature of the emotion fear, the boundary of one intensity level mixes in the boundary of another intensity level. Hence, the intensity levels lying between 0.24 and 0.5 will be interpreted as low undesirability and lower than these values as the very low undesirability. In the same way, the other intensity levels of undesirability variable can be interpreted. In the same way, Table 5 and Table 6 are showing the five quantitative values for finding the different intensity levels of likelihood and Ig variables. These values are then provided to the EIC_Agent for computing different intensities of fear.

Table 4 Quantitative Values of Five Intensity levels of Desirable Variable

| **VLD** | **LD** | **MD** | **HD** | **VHD** |
| --- | --- | --- | --- | --- |
| 0-0.24 | 0.1-0.5 | 0.25-0.73 | 0.51-0.9 | 0.76-1 |

Table 5 Quantitative Values of Five Intensity levels of Likelihood Variable

| VLL | LL | ML | HL | VHL |
|---|---|---|---|---|
| 0-0.24 | 0.1-0.5 | 0.25-0.73 | 0.51-0.9 | 0.76-1 |

Table 6 Quantitative Values of Five Intensity levels of Global Variable (Ig)

| VLIg | LIg | MIg | HIg | VIg |
|---|---|---|---|---|
| 0-0.24 | 0.1-0.5 | 0.25-0.73 | 0.51-0.9 | 0.76-1 |

## 7.2 Experiment 2

This section presents the results and discussion of validation experiments. First the results and discussion of Invariant 1 tests have been presented. Then the results and discussion of invariant 2 tests have been presented. In the last the results and discussion of invariant 3 have been presented.

### 7.2.1 Invariant 1

The graphical results of invariant1 experiments have been presented in figure 14. The x-axis of the graph is showing the distance and fear intensity, whereas the vertical axis is presenting the number of samples. If we examine the first sample then it can be seen that when the distance between AV and BSSP is 15 patches, each patch = 5 meters in the simulation, then the fear perceived by EIC_Agent is 0, means no fear at all. In a second sample, it can be seen that the distance between AV and BSSP reduced to the 13.65 patches and the fear intensity level of EIC_Agent increased from 0 to 0.1 i.e. very low fear. In the same way. If we examine the fourth sample then the distance between AV and BSSP is 9.66 patches and the corresponding fear level is 0.4 which is high low fear. In the very next sample, the distance reduced to 7.66 patches and EIC_Agent perceived medium fear i.e. 0.5. Further, if we examine the sample 7 and 8 then it can be seen that when the distance between AV is 2.66 and 1.66 patches then the fear intensity is 2.66 0.69 and 0.8 respectively, whereas 0.69 means high fear and 0.8 means very high fear. From these results, it can be validated that EIC_Agent exhibits fear intensity level according to the distance between itself and the BSSP. If the distance is high, medium or low then the fear intensity level also lies in high, medium or low range.

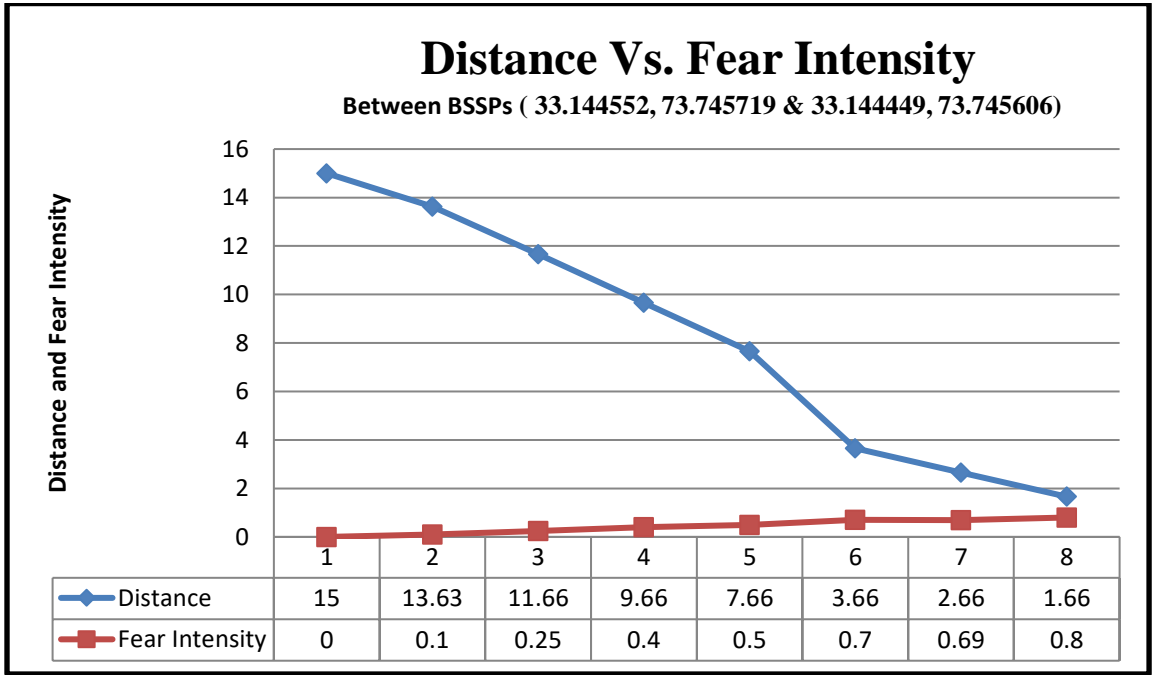

Fig. 14 Spectrum Mobility Mechanism of EIC_Agent Using Fear based PDFA

## 7.2.2 Invariant 2

Figure 15 is presenting the results of invariant 2 experiments. The aim of these experiments was to prove that when the future signal strength of whitespace, in-use, is getting worse, then the EIC_Agent extended CR-Site will select the new white space with good signal strength. From the figure 15. It can be seen that when the simulation was started, then EIC_Agent noted that current white space in use is Telenor, its current signal strength is -91 dBm and its future signal strength will be -70 dBm. Furthermore, EIC_Agent consulted its database and found that white space of the Zong service provider is available with the signal strength of -50 dBm. Hence, it can be seen that EIC_Agent performed spectrum mobility, to the Zong whitespace, by controlling sensing, whitespace optimization and shifting modules of CR-Site. After that, EIC_Agent noted that the future signal strength of current white space, Zong, will be -70 dBm means the quality of communication can be affected then it consulted its database and found Telenor, having signal strength of -29 dBm, as the best option and performed spectrum mobility from Zong to the Telenor by controlling the CR-Site functions under the influence of different fear levels. If we examine the fourth experiment, which is different than previous spectrum handovers, then it can be seen that EIC_Agent extended CR-Site was using Ufone white space with signal strength of -45 dBm and keep using it on the next BSSP instead of clear degradation in signal strength (from -45 dBm to -65 dBm) because no other best whitespace, other than Ufone was available. In the continuation of the experiment, if we further analyse the results, then it can be seen that the EIC_Agent extended CR-Site select new white space with good signal strength, whenever the future signal strength of whitespace, in-use, is getting bad. Hence, these results further validate that the proposed spectrum mobility mechanism of EIC_Agent extended CR-Site is working properly.

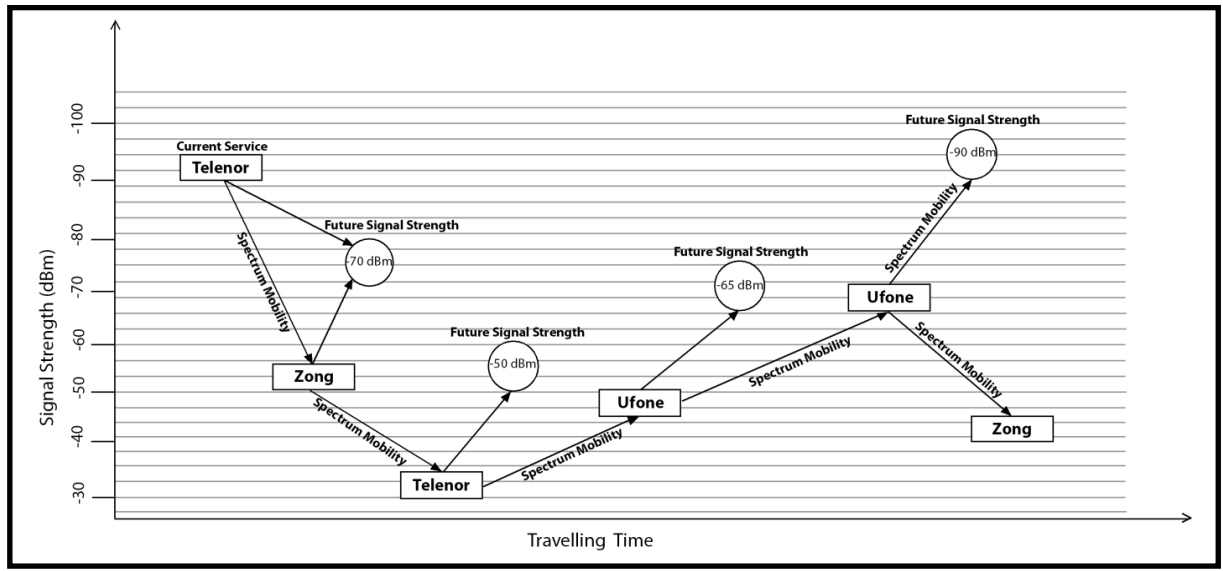

Fig. 15 Spectrum Mobility Mechanism of EIC_Agent Using Fear based PDFA

### 7.2.3 Invariant 3

The results of invariant 3 experiments have been presented in table 7, 8 and 9. First, let us examine the results of table 7, which is presenting the worst case, when the spectrum sensing time =200 ms, TCP connection setup time = 5 seconds and white space optimization time taken by MEGA is 0.527 microseconds. From the test 1, it can be seen that the EIC_Agent failed to perform spectrum mobility because time left to the reach the BSSP is 1.25 seconds and the time required to perform the spectrum mobility is 5.201 seconds. In the second experiment, the EIC_Agent extended CR-Site performed the spectrum mobility scheme successfully as the time left to reach the BSSP is greater than the required spectrum mobility time. If we analyze the overall results, then it can be seen that EIC_Agent extended CR-Site performed four time spectrum mobility task successfully, whereas it failed for the 6 times. The graphical representation of table 7 is presented in figure 16a. Now, if we analyze the results of table 8, which is presenting the average case, when the spectrum sensing time =100 ms, TCP connection setup time = 2 seconds and white space optimization time taken by MEGA is 0.527 microseconds, then it can be seen that EIC_Agent extended CR-Site has performed nine (9) successful spectrum handovers as compared to the worst case scenario. The graphical representation of table 8 is presented in figure 16b. In the last, if we examine the results of table 9, which is presenting the best case, when the spectrum sensing time =50 ms, TCP connection setup time = 1 second and white space optimization time taken by MEGA is 0.527 microseconds, then EIC_Agent extended CR-Site has performed 100 % spectrum mobility tasks without any failure. The graphical representation of table 9 is presented in figure 16c. From all of this discussion, it can be seen that the performance of the proposed spectrum mobility scheme of EIC_Agent extended CR-Site will be best when the TCP connection setup time will be less.

TABLE 7 STATE TRANSITION TABLE OF PDFA USING FEAR FACTOR

| Distance from BSSP | Fear Intensity | Required Spectrum Mobility Time | Time Left to Reach the BSSP |
|---|---|---|---|
| 1 | 0.860991246 | 5.201 | 1.25 |
| 9 | 0.036175583 | 5.201 | 11.25 |
| 13 | 0.067837111 | 5.201 | 16.25 |
| 3 | 0.668127954 | 5.201 | 3.75 |
| 2 | 0.855923088 | 5.201 | 2.5 |
| 5 | 0.466675244 | 5.201 | 6.25 |
| 3 | 0.674727595 | 5.201 | 3.75 |
| 2 | 0.837151607 | 5.201 | 2.5 |
| 10 | 0.245635263 | 5.201 | 12.5 |
| 4 | 0.465293116 | 5.201 | 5 |

TABLE 8 STATE TRANSITION TABLE OF PDFA USING FEAR FACTOR

| Distance from BSSP | Fear Intensity | Required Spectrum Mobility Time | Time Left to Reach the BSSP |
|---|---|---|---|
| 1 | 0.86 | 2.1 | 1.25 |
| 9 | 0.04 | 2.1 | 11.25 |
| 13 | 0.07 | 2.1 | 16.25 |
| 3 | 0.67 | 2.1 | 3.75 |
| 2 | 0.86 | 2.1 | 2.5 |
| 5 | 0.47 | 2.1 | 6.25 |
| 3 | 0.67 | 2.1 | 3.75 |
| 2 | 0.84 | 2.1 | 2.5 |
| 10 | 0.25 | 2.1 | 12.5 |
| 4 | 0.47 | 2.1 | 5 |

TABLE 9 STATE TRANSITION TABLE OF PDFA USING FEAR FACTOR

| Distance from BSSP | Fear Intensity | Required Spectrum Mobility Time | Time Left to Reach the BSSP |
|---|---|---|---|
| 1 | 0.860991246 | 1.05 | 1.25 |
| 9 | 0.036175583 | 1.05 | 11.25 |
| 13 | 0.067837111 | 1.05 | 16.25 |
| 3 | 0.668127954 | 1.05 | 3.75 |
| 2 | 0.855923088 | 1.05 | 2.5 |
| 5 | 0.466675244 | 1.05 | 6.25 |
| 3 | 0.674727595 | 1.05 | 3.75 |
| 2 | 0.837151607 | 1.05 | 2.5 |
| 10 | 0.245635263 | 1.05 | 12.5 |
| 4 | 0.465293116 | 1.05 | 5 |

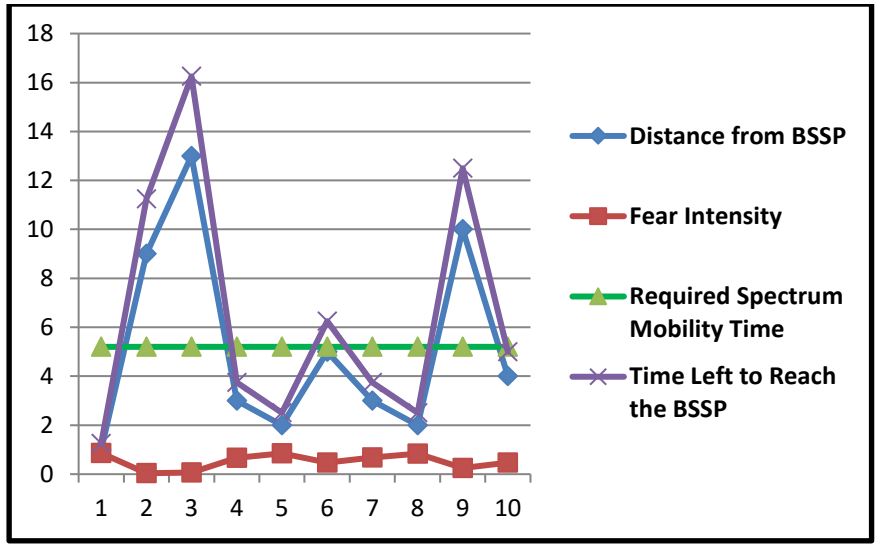

(a)

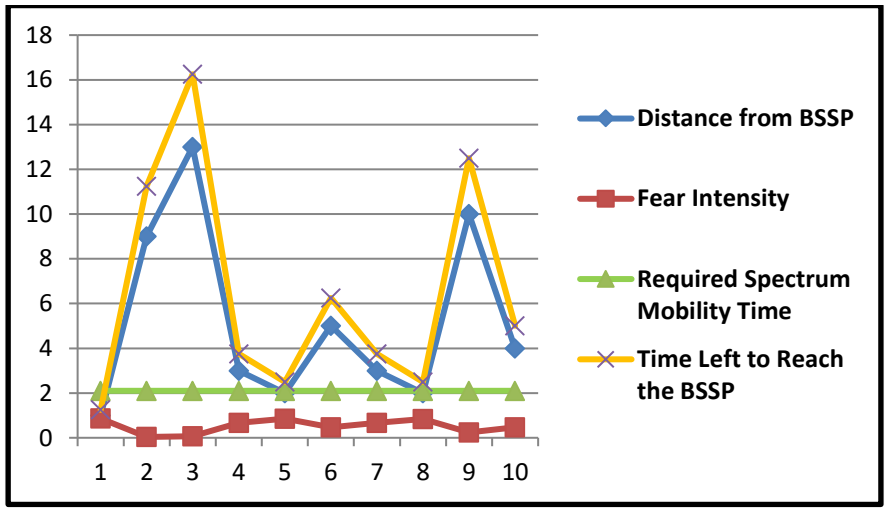

(b)

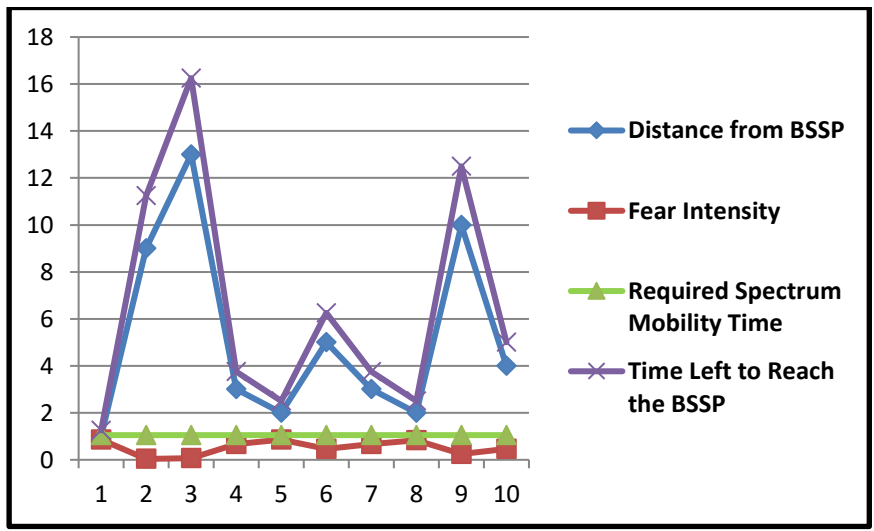

(c)

Fig. 16 Graphical representation of invariant 3 results (a) Spectrum sensing time =200 ms, TCP connection setup time = 5 seconds and white space optimization time taken by MEGA =0.527 microseconds, (b) Spectrum sensing time =100 ms, TCP connection setup time = 2 seconds and white space optimization time taken by MEGA =0.527 microseconds, (c) Spectrum sensing time =50 ms, TCP connection setup time = 1 second and white space optimization time taken by MEGA = 0.527 microseconds

## 8. Qualitative Comparison with the Existing State of the Art

In the existing literature, different types of handover schemes for cognitive radio and non-cognitive radio based wireless networks have been proposed. We have compared our proposed spectrum mobility scheme with both of these schemes.

At present, there is no research available, other than our own previous research work [39], for spectrum handover schemes in CR vehicular networks using human emotions approach. Furthermore, to the best of our knowledge and research databases search, we have found only three recently published research works by a same author Kumar et. al [11-13], which are addressing the spectrum mobility problem in cognitive radio based vehicular networks. First one is an optimal network selection scheme for the spectrum handover decision in NEMO (Network Mobility) based CR vehicular networks [11]. The authors have addressed the problem of the multi-mode vehicle, which is connected with multiple radio access technologies and has the capability to call more than one non-safety applications like voice, video and best effort services. In such scenario, optimal network selection for handover decision is a complex problem for multi-mode vehicles. To address this issue, authors have utilised grey relational analysis (GRA) and cost based method of multiple attributes decision making (MADM) technique. Furthermore, the velocity of the vehicle has been used as a decision parameter to initiate the spectrum handover between the different networks. Second one is another optimal network selection scheme using game theoretic auction theory for efficient spectrum handover in cognitive radio based vehicular networks [12]. The authors have modelled the vehicular networks as tenderee /Buyer and available networks as seller/bidder entities. Third one is a context-aware spectrum handover scheme with multiple attributes decision-making (MADM) methods for preferred network selection in CR vehicular networks [13].

However, the main focus of these three research works is finding the optimal whitespaces for efficient spectrum handover. The authors have not discussed the proactive timeline in any of these research works, which answers that when and where to perform the handover measurement & initiation mechanism, which help in initiating a handover towards a new network, handover decision measurement, which help in comparing the results with predefined values to decide whether to perform the handover or not and handover execution mechanism, which define the connection procedure with the new network. In other sense, these proposed handover schemes are just considering one small aspect of the overall handover procedure. In contrast to this research work, we have employed the CR-Site [7], which helps in selecting optimal network selection or optimal whitespaces using MEGA and further improved the spectrum mobility mechanism of CR-Site with the help of emotions inspired approach. We have addressed all three phases of spectrum handover which are handover measurement & initiation mechanism, handover decision measurement and handover execution procedure [18]. For handover measurement & initiation mechanism, we have utilised the concept of pre-build BSSPs database. Using this BSSP's database EIC_Agent performs the measurements of current RSS of white space in use. To find out that is there a need to initiate the spectrum mobility procedure or not. Then EIC_Agent compares the results of measurement with predefined values and decide is there a need for performing the handover or not. Then the handover execution process is executed with the help of the shift module of CR-Site.

TABLE 10 TABULAR COMPARISON OF PROPOSED SCEHME WITH STATE-OF-THE-ART ADDRESSING SPECTRUM HANDOVER IN COGNITIVE RADIO BASED VEHICULAR NETWORKS

| S.No | Author | Contribution | Type of Applications | Address Handover Measurement & Initiation Mechanism | Address Handover Decision | Address Handover Execution Mechanism |
|---|---|---|---|---|---|---|
| 1 | Kumar et. al [11] | An optimal network selection scheme for the spectrum handover decision in NEMO (Network Mobility) based CR vehicular networks | Non-Safety | No | Yes | No |
| 2 | Kumar et. al [12] | An optimal network selection scheme using game theoretic auction theory for efficient spectrum handover in cognitive radio based vehicular networks | Non-Safety | No | Yes | No |
| 3 | Kumar et. al [13] | An optimal network selection scheme using context-aware approach with multiple attributes decision-making (MADM) method | Non-Safety | No | Yes | No |
| 4 | Our proposed Scheme | Fear inspired pure proactive spectrum mobility scheme in cognitive radio based vehicular netwroks, which is addressing all three phases of handover process | Safety | Yes (RSS and Distance between BSSPs) based | Yes (RSS Based) | Yes |

Furthermore, there is extensive research work available for general handover schemes in cognitive radio based wireless networks. A fuzzy logic inspired channel selection procedure for efficient handover in cognitive radio based networks has been proposed by Salgado et al. [16]. The authors have utilised multiple-criteria decision-making approach to select the optimal channel selection in prior, which can be utilised by the CR-node to perform the channel handover. However, the authors have not discussed handover measurement & initiation and handover execution mechanism. In another research work, Ahmed et al. [17] have proposed handover decision-making process using PU activity and SINR as handover decision parameters along with optimal channel selection. For this purpose, they have proposed a fuzzy logic inspired scheme with 18 fuzzy rules, which help in making spectrum handover decision and computing the channel gain to select the optimal channel. The drawback of this paper is that it does not discuss any handover execution mechanism. Potdar and Patil [14] has utilised fuzzy logic and neural network to propose an efficient spectrum handover scheme for CR. The authors focused on the mobility of secondary user (SU) with intracellular and intercellular handovers to provide quality of service. Furthermore, they have used a novel "Resource Usability Parameter" to prioritise handover in critical situations. Human emotions and fuzzy logic inspired proactive spectrum mobility scheme has been proposed by Riaz and Niazi [39]. The authors have considered RSS and distance as main parameters of spectrum mobility in General Packet Radio Service (GPRS) based vehicular networks. In addition to these state of the art handover schemes for cognitive radio based networks, there exist non-cognitive radio schemes as well which utilised the Received Signal Strength (RSS) as we have utilised RSS as spectrum decision parameter. Few of them have been discussed as under. For example, Saxena and Roy [40] has proposed a proactive handover scheme considering Received Signal Strength (RSS) as spectrum mobility parameter to improve the performance of handovers in WLANs. In [41], Mohanty et al. have proposed RSS based spectrum handover protocol to perform spectrum switching between WLAN and 3G networks. In the same way RSS parameter has been utilised to propose an efficient handover production scheme by Becwar et al. [42].

However, the problem with these fuzzy logic based handover schemes [14][16-17][44-45] and [47] for cognitive radio and non-cognitive radio based wireless networks is that the process of fuzzification and defuzzification involved in decision processing causes a delay in the handover procedure. Furthermore, the fuzzy logic based handover schemes rely on the number of fuzzy rules, an excessive number of such will straightforwardly prejudice its efficiency [43]. In addition, with the increase in the fuzzy subsets, the fuzzy rules increases exponentially, which increases the computation time of handover decision making. In contrast to these fuzzy logic rules, we have proposed just four fear intensity inspired rules, which control the handover scheme from handover measurement & initiation to handover decision and handover decision to handover execution. Furthermore, the problem with these solutions is that they are giving solutions for making handover decisions in the heterogeneous wireless environment, whereas we have addressed the problem where the wireless environment is homogeneous i.e. GSM networks but the service providers are heterogeneous such as Ufone, Telenor, Mobilink, and Zong. In addition, all of the above-mentioned handover techniques do not provide the complete solution from Handover Measurement Initiation to Handover Execution. Whereas, we have proposed proactive spectrum mobility scheme that helps the AV to complete all of the three stages of spectrum handover intime and successfully.

TABLE 11 Tabular Comparison of Proposed Scehme With State-of-the-art addressing spectrum handover Using Fuzzy Logic

| S.No | Author | Contribution | Approach Used | Number of controlling rules | Address Cognitive Radio Based Vehicular Networks |
|---|---|---|---|---|---|
| 1 | Ahmed et al. [17] | Proposed Fuzzy logic based handover decision-making process using PU activity and SINR as handover decision parameters | Fuzzy Logic | 18 | No |
| 2 | Potdar and Patil [14] | Proposed fuzzy logic and neural network based efficient spectrum handover scheme for CR networks | Fuzzy Logic/Neural Network | 27 | No |
| 3 | Riaz and Niazi [39] | Proposed fuzzy logic based fear inspired spectrum mobility scheme | Fuzzy Logic | 25 | Yes |
| 3 | Our proposed Scheme | Proposed a Fear inspired pure proactive spectrum mobility scheme in cognitive radio based vehicular networks with less number of controlling rules | Fear Emotion | 4 | Yes |

# 9. Conclusion

In-time spectrum mobility for syntactic interoperability in Cognitive Radio (CR) based high-speed IoV systems is a considerable challenge due to the unpredictable radio frequency environment. This is especially for use in lossless data communication, while no such work has been reported previously to solve this problem in vehicular networks. In this paper, we propose a novel pure proactive spectrum mobility scheme using fear emotion. A probabilistic deterministic finite automaton using fear factor is proposed to perform efficient spectrum mobility using fuzzy logic. The system was tested using active data from different GSM service providers on Mangla-Mirpur road. As a future, the GSM network related database can be replaced with 4G networks signal strength data and the same scheme can be utilized to propose a highly reliable syntactic interoperable mechanism by performing in-time spectrum handoffs in high speed IoV networks.